\documentclass{PoS}

\title{Heavy quark correlations \\in hadronic collisions}

\ShortTitle{Heavy quark correlations}

\author{\speaker{Marta {\L}uszczak}\\
        University of Rzesz\'ow, ul. Rejtana 16c\\
        PL-35-959 Rzesz\'ow, Poland\\
        E-mail: \email{luszczak@univ.rzeszow.pl}}

\author{Antoni Szczurek\\
        Institute of Nuclear Physics, PL-31-342 Cracow, Poland\\
        and University of Rzesz\'ow, PL-35-959 Rzesz\'ow, Poland\\
        E-mail: \email{Antoni.Szczurek@ifj.edu.pl}}

\abstract{ We discuss results of the $k_t$ - factorization approach
for heavy quark-heavy antiquark correlations 
in proton-proton and proton-antiproton collisions for RHIC,
Tevatron and LHC. We consider correlations in the azimuthal angle 
as well as in the two-dimensional space 
of transverse momentum of heavy quark and heavy antiquark.
We compare results obtained with the help 
of different unintegrated parton distributions (UPDF) from the literature.}

\FullConference{High-pT physics at LHC\\
		 March 23-27 2007\\
		 University of Jyv\"askyl\"a, Jyv\"askyl\"a, Finland}

\begin{document}

%---------------------
\section{Introduction}
%---------------------

The heavy quark-antiquark hadroproduction
is known as very important test of conventional 
gluon distributions within a standard factorization approach.
Standard collinear approach does not include transverse momenta of initial gluons.
The method to include transverse momenta is $k_t$ - factorization approach \cite{CCH91,CE91,BE01}.
At the leading order of the collinear approach the heavy quark and heavy antiquark 
are produced back-to-back. In the unintegrated parton distributions (UPDF) 
approach \cite{KL01,GBW_glue,KMR}, the azimuthal angle and $p_t$ decorrelations
(from the collinear leading-order configurations) are obtained 
already in the leading order of perturbative expansion    
\cite{LS06}. In Ref.\cite{LS06} we have explored in detail $c \bar c$
kinematical correlations at the Tevatron energies. Here we wish to
discuss a more general case of (heavy quark)--(heavy antiquark) correlations for the RHIC,
Tevatron and LHC energy range. We wish to empasize that this is not yet
well explored field of high-energy physics which could and should be studied
in the future, in particular at LHC.

%--------------------------------
\section{(Heavy quark)-(heavy antiquark) correlatons}
%--------------------------------

%-------------------------------
\begin{figure}[!htb] % Figure 1
\begin{center}
\includegraphics[width=6.2cm]{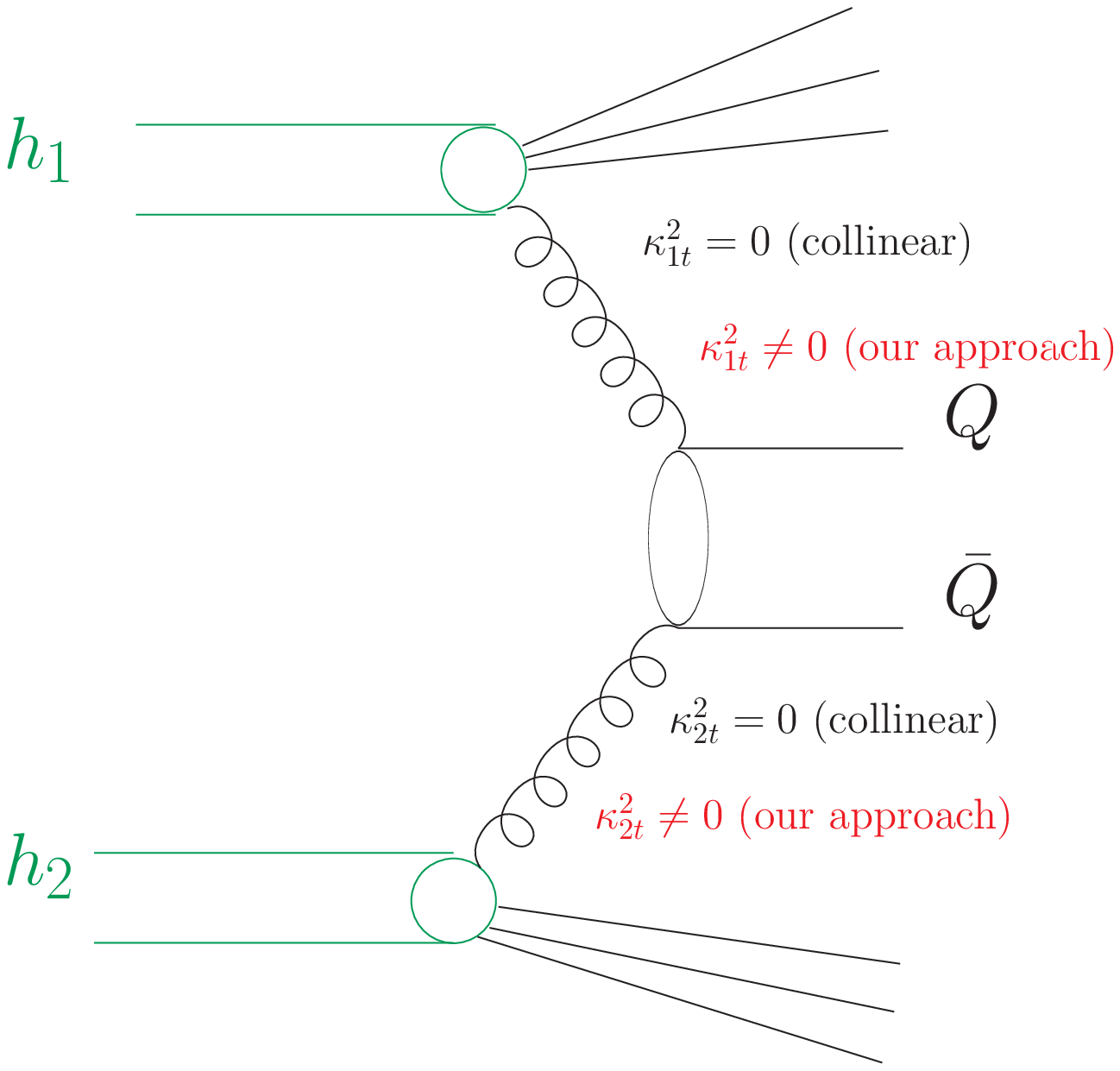}
\includegraphics[width=6.0cm]{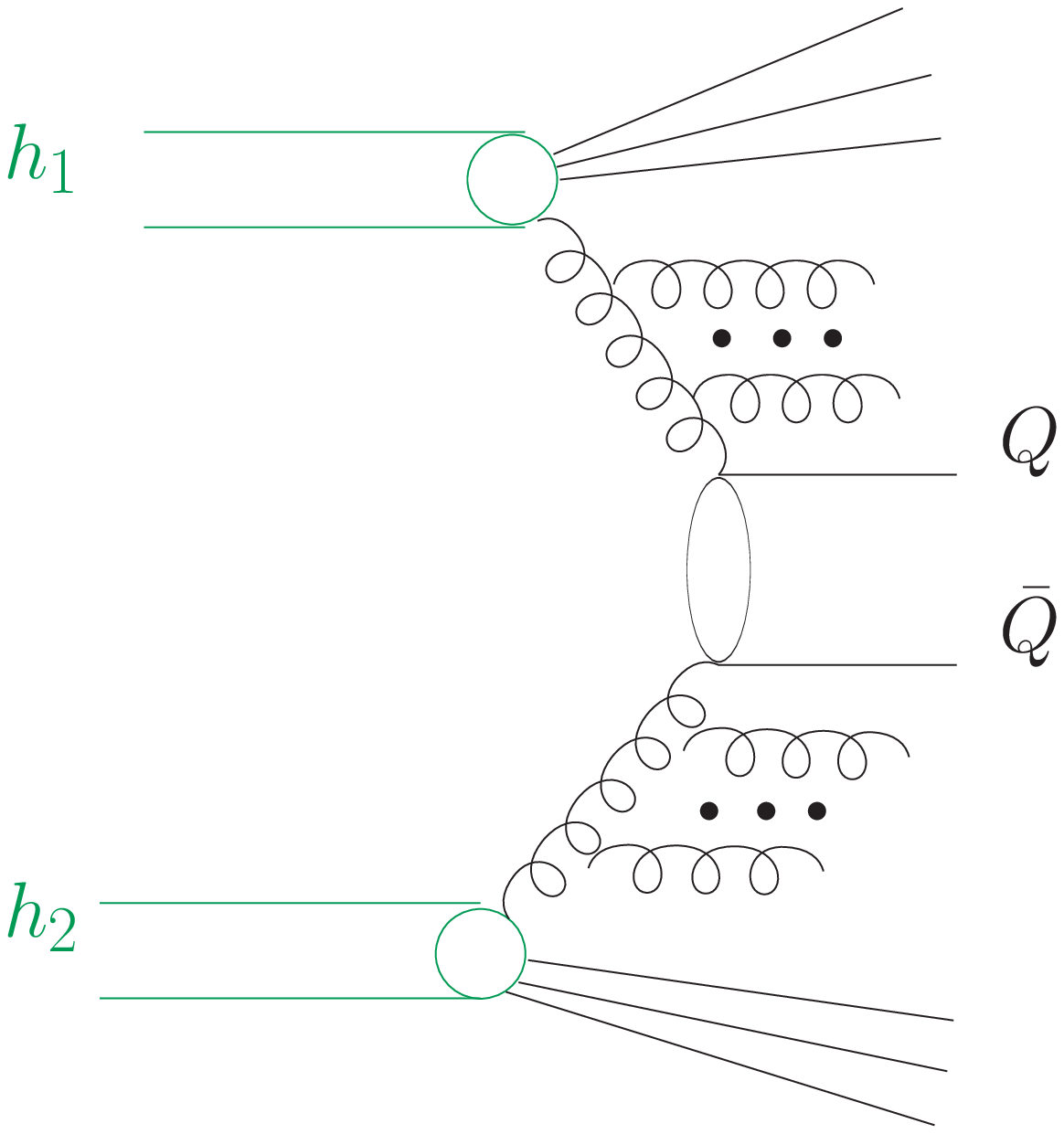}
\caption[*]{
Typical diagrams for heavy quark - antiquark production.
\label{fig:d_gluon}
}
\end{center}
\end{figure}
%-----------------------------------

Let us consider the reaction $h_1 + h_2 \to Q + \bar Q + X$ (see Fig.\ref{fig:d_gluon}),
where $Q$ and $\bar Q$ are heavy quark and heavy antiquark,
respectively.
In the leading-order $k_t$- factorization approach the cross section in rapidity of $Q$ ($y_1$),
in rapidity of $\bar Q$ ($y_2$) and transverse momentum of heavy quark
 ($p_{1,t}$) and heavy antiquark ($p_{2,t}$) can be written as
\begin{eqnarray}
\frac{d \sigma}{d y_1 d y_2 d^2p_{1,t} d^2p_{2,t}} = \sum_{i,j} \;
\int \frac{d^2 \kappa_{1,t}}{\pi} \frac{d^2 \kappa_{2,t}}{\pi}
\frac{1}{16 \pi^2 (x_1 x_2 s)^2} \; \overline{ | {\cal M}_{ij} |^2}
\nonumber \\  
\delta^{2} \left( \vec{\kappa}_{1,t} + \vec{\kappa}_{2,t} 
                 - \vec{p}_{1,t} - \vec{p}_{2,t} \right) \;
f_i(x_1,\kappa_{1,t}^2) \; f_j(x_2,\kappa_{2,t}^2) \; ,
\label{LO_kt-factorization}    
\end{eqnarray}
where $f_i(x_1,\kappa_{1,t}^2)$ and $f_j(x_2,\kappa_{2,t}^2)$
are so-called unintegrated parton distributions.
The two extra factors $1/\pi$ attached to the integration over
$d^2 \kappa_{1,t}$ and $d^2 \kappa_{2,t}$ are due to the conventional
relation between unintegrated and integrated parton distributions. 
The two-dimensional delta function assures momentum conservation.

The unintegrated parton distributions must be evaluated at:
$x_1 = \frac{m_{1,t}}{\sqrt{s}}\left( \exp( y_1) + \exp( y_2) \right)$,
$x_2 = \frac{m_{2,t}}{\sqrt{s}}\left( \exp(-y_1) + \exp(-y_2) \right)$.
The matrix element must be calculated for initial
off-shell partons. The corresponding formulae for initial gluons
were calculated in \cite{CCH91,CE91} (see also \cite{BE01}).
In our paper  we compare results
obtained for both on-shell and off-shell matrix elements for charm-anticharm correlations.

%----------------
\section{Results}
%----------------

%--------------------------------------------------------------------
\begin{figure}[!thb] % Figure 2
\begin{center}
\includegraphics[width=7cm]{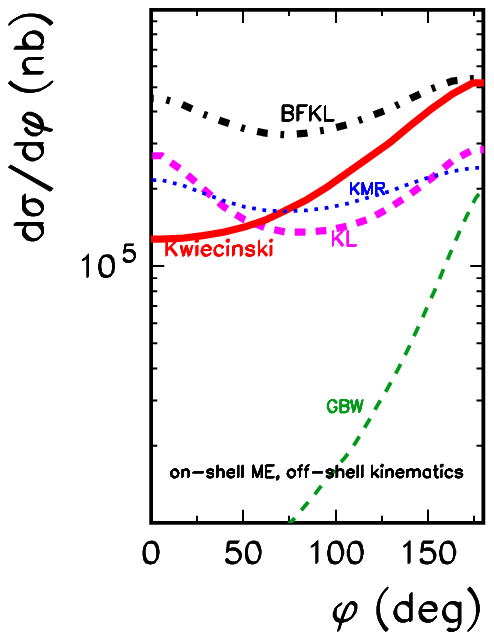}
\includegraphics[width=7cm]{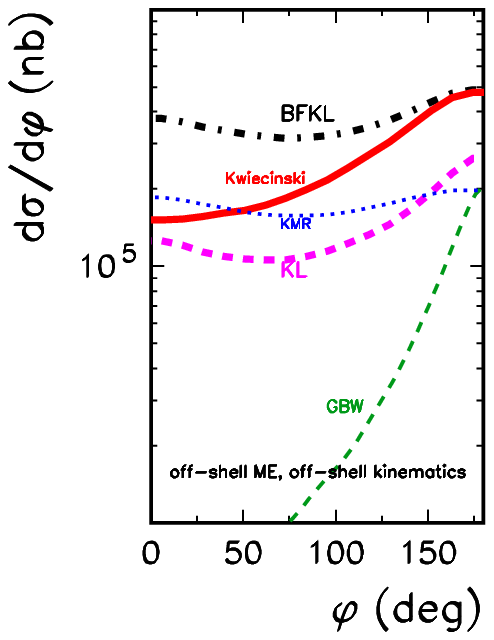}
\caption[*]{
$d\sigma/d\phi$ for charm - anticharm production at  W = 1960 GeV
for different UGDFs.
The results with on-shell kinematics are shown in the left panel
and results with off-shell kinematics in the right panel.
In this calculation both factorization and renormalization scales
were fixed for 4 $m_c^2$.
\label{fig:dsig_dpt_gauss2}
}
\end{center}
\end{figure}

%--------------------------------------------------------------------
In Fig.2  we compare results for
different unintegrated gluon distributions from the literature.
Different results are obtained for different UGDF. BFKL dynamics 
leads to strong decorrelations in azimuthal angle between charm and anticharm quarks. 
In contrast, the nonperturbative GBW glue leads to strong azimuthal correlations between $c$
and $ \bar c$. The saturation idea inspired KL distribution leads to an local
enhancement for $\phi_{c \bar c} \approx$ 0 which is probably due
to simplifications made in parametrizing the KL UGDF.
In the last case there is sizeable difference between the result
obtained with on-shell (left panel) and off-shell (right panel)
matrix elements. All this is due to an interplay of the matrix element 
and the unintegrated gluon distributions.

%--------------------------------------------------------------------

\begin{figure}[!thb] % Figure 3
\begin{center}
    \includegraphics[height=6.8cm]{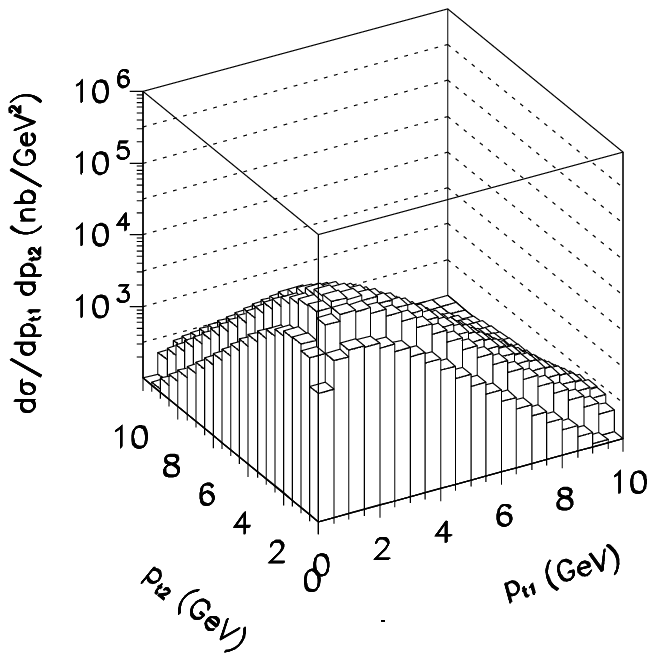}
    \includegraphics[height=6.8cm]{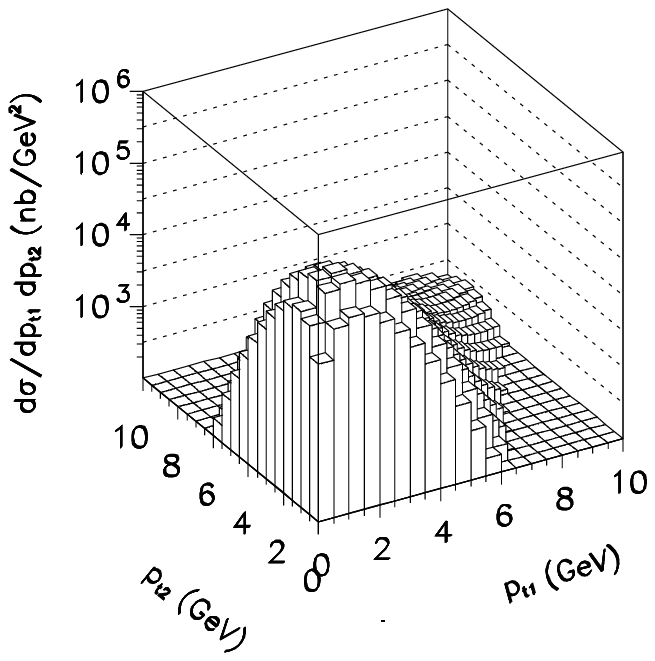}
    \includegraphics[height=6.8cm]{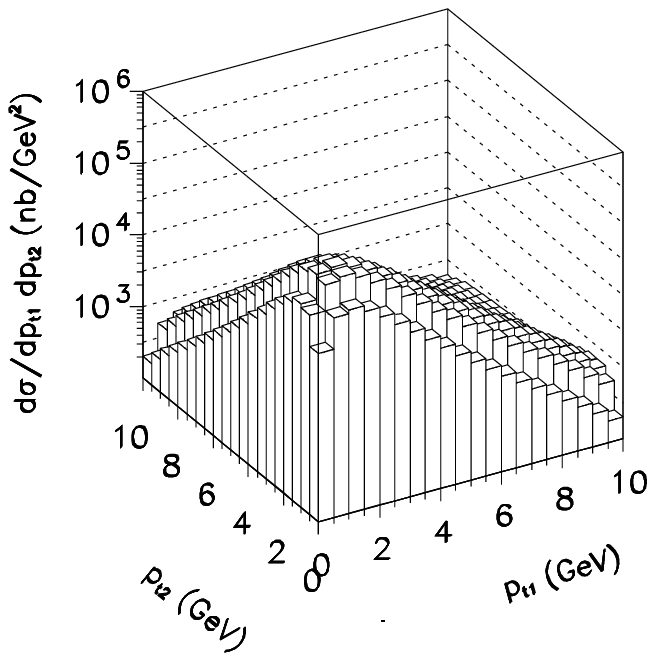}
    \includegraphics[height=6.8cm]{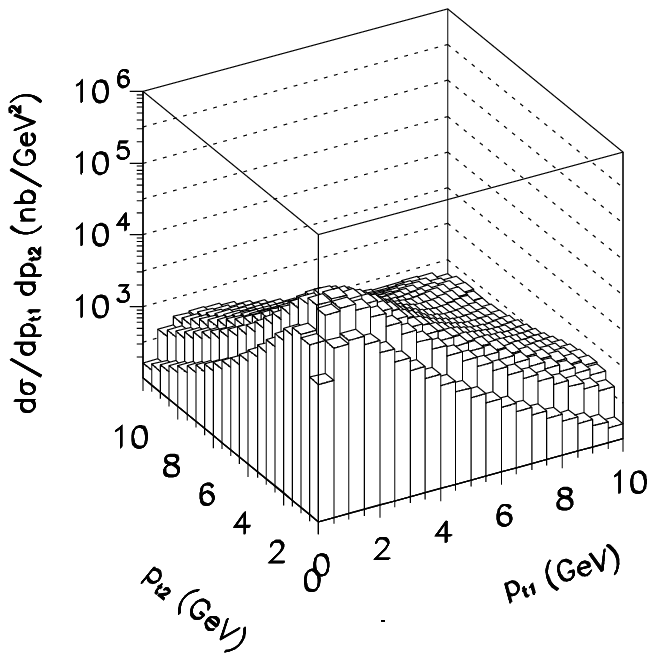}
\caption[*]{
Two dimensional distributions in ($p_{1,t}$) of charm quark and ($p_{2,t}$) of charm antiquark for KMR, 
Kwieci\'nski ($b_0=1$~GeV$^{-1}$, $\mu^2=10$~GeV$^2$), BFKL, KL UGDFs  for W= 1960 GeV.
\label{fig:dsig_dpt_gauss3}
}
\end{center}
\end{figure}
%------------------------------
\begin{figure}[!thb] % Figure 4
\begin{center}
    \includegraphics[height=6.8cm]{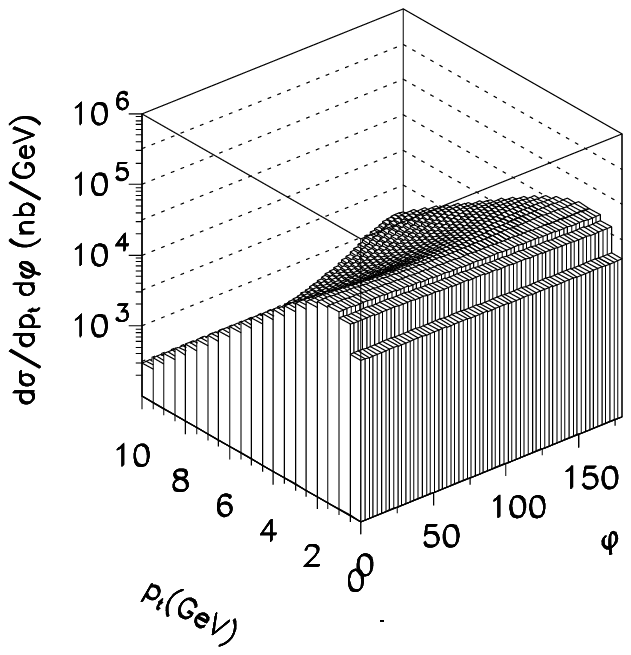}
    \includegraphics[height=6.8cm]{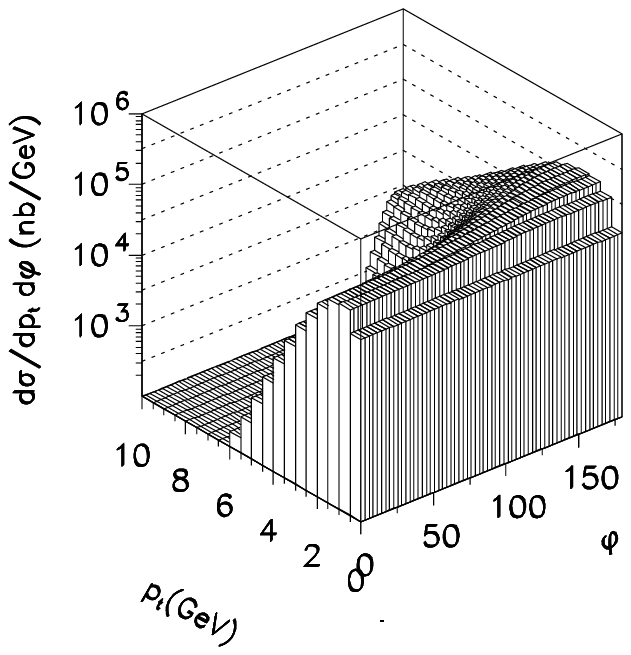}
    \includegraphics[height=6.8cm]{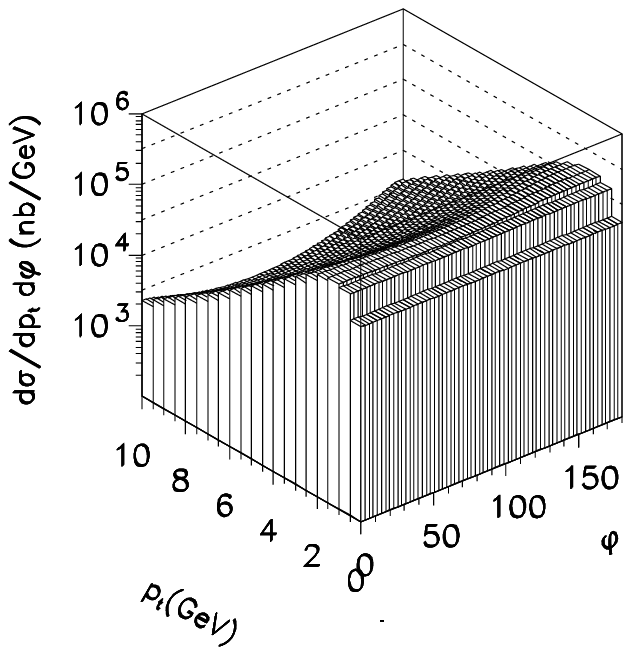}
    \includegraphics[height=6.8cm]{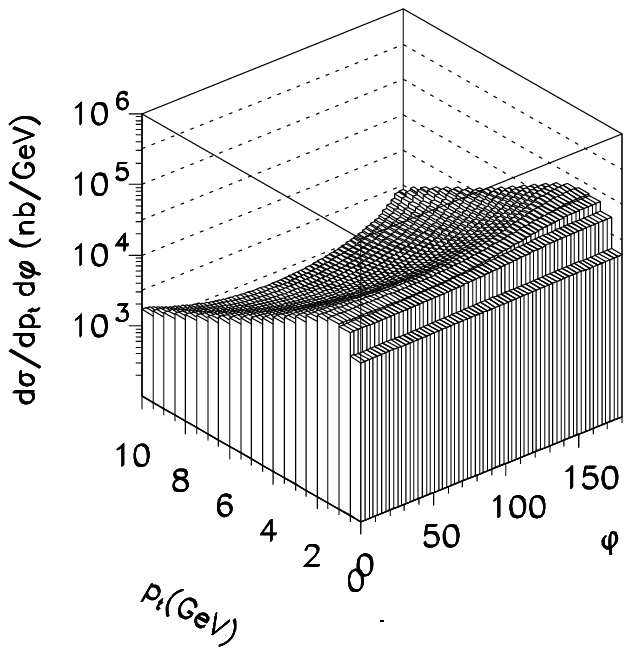}
\caption[*]{
Two dimensional distributions in $p_t$ and $\phi$ for KMR, 
Kwieci\'nski ($b_0=1$~GeV$^{-1}$, $\mu^2=10$~GeV$^2$), BFKL, KL for W = 1960 GeV.
\label{fig:dsig_dpt_gauss4}
}
\end{center}
\end{figure}
 
In Fig.3 we show results for correlations in ($p_{1,t}$) of $c$ and ($p_{2,t}$) of
$\bar c$ for different unintegrated gluon distributions. Our results depend
strongly on different UGDF. This is very interesting, because it can be verified
in future experimental studies.

\begin{figure}[!thb] % Figure 5
\begin{center}
\includegraphics[width=7cm]{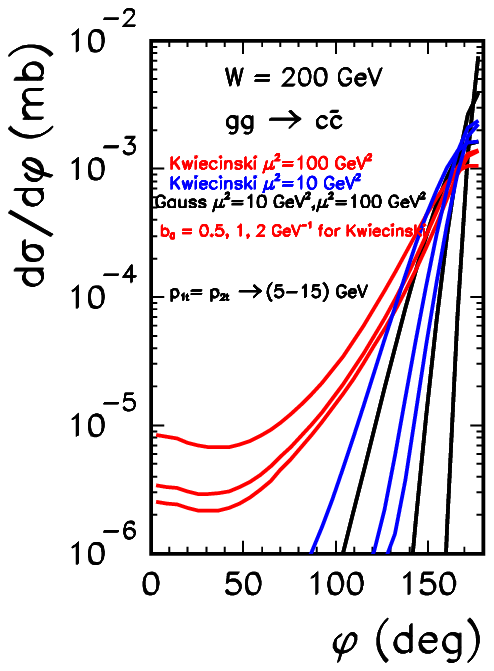}
\includegraphics[width=7cm]{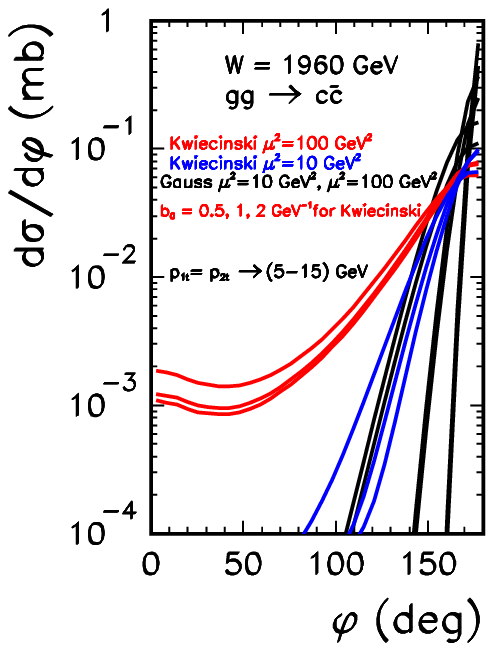}
\caption[*]{
(Color on-line) The azimuthal correlations for the gg $\to c \bar c$ for RHIC 
and Tevatron energy with the Kwieci\'nski UGDFs for different values 
of the nonperturbative parameter $b_0$ ($0,5$~GeV$^{-1}$, $1$~GeV$^{-1}$, $2$~GeV$^{-1}$) 
and for different evolution scales $\mu^2=10$~GeV$^2$ (blue line) 
and $\mu^2=100$~GeV$^2$ (red line). The initial distributions ( without evolution) 
are shown for reference by black lines.
\label{fig:dsig_dpt_gauss5}
}
\end{center}
\end{figure}

In Fig.4 we show another type of two-dimensional distribution.
Generally, the bigger transverse momentum of the produced quark (or antiquark)
the stronger back-to-back correlation in azimuth. Similarly as in the previous
case, the results depend on UGDFs used in the calculation. The somewhat stronger
back-to-back correlation in the case of the Kwieci\'nski UGDF requires a separate discussion. 
As an example in Fig.5 we show the dependence of the azimuthal correlations 
for $p_{1,t}, p_{2,t} \in$ (5,15) GeV. We show the result for different values 
of the parameter $b_0$ describing initial (for QCD evolution)
nonperturbative distribution of gluons in nucleons as well as for different
values of the scale parameter $\mu^2$. The results can be characterized
as follows. The smaller $b_0$, the bigger the decorrelation between $c$ and
$\bar c$. On the other hand, the decorrelation increases with raising $\mu^2$.
Physically one would expect $\mu^2 = \mu^2(p_{1,t},p_{2,t})$.
A reasonable choice would be
\begin{equation}
\mu^2(p_{1,t},p_{2,t}) = C_1 m_c^2 + C_2 (p_{1,t}^2 + p_{2,t}^2) \; ,
\label{running_scale}
\end{equation}
where $C_1, C_2 \sim$ 1 may be expected. This would require a use
of running scale instead a fixed one as at present. In the moment
this was not possible technically and therefore only typical values
of the scales were taken (see Fig.\ref{fig:dsig_dpt_gauss5}). The scale $\mu^2$ = 100 GeV$^2$
is justified by the averaged values of the momenta in our calculation
$\langle p_{1,t}^2 \rangle , \langle p_{2,t}^2 \rangle \approx$ 100 GeV$^2$.
The figure shows that QCD evolution embedded in the Kwieci\'nski equations
leads to a strong extra decorrelation of $c$ and $\bar c$ in addition
to that of the nonperturbative origin as encoded in the parameter $b_0$.

%--------------------------------------------------------------------

%The smaller $b_0$ the bigger decorrelation  in azimuthal angle can be observed.
%In fig.7 we show also the role of the evolution scale in the Kwieci\'nski %distributions. 

\begin{figure}[!thb] % Figure 6
\begin{center}
\includegraphics[width=7cm]{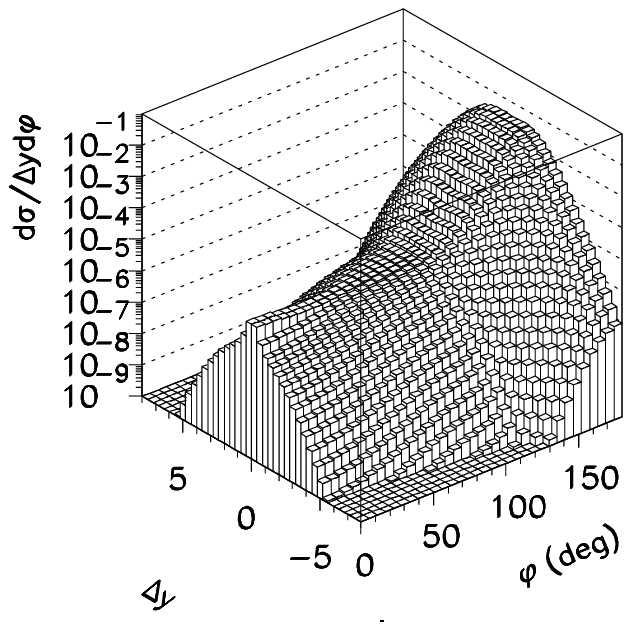}
\includegraphics[width=7cm]{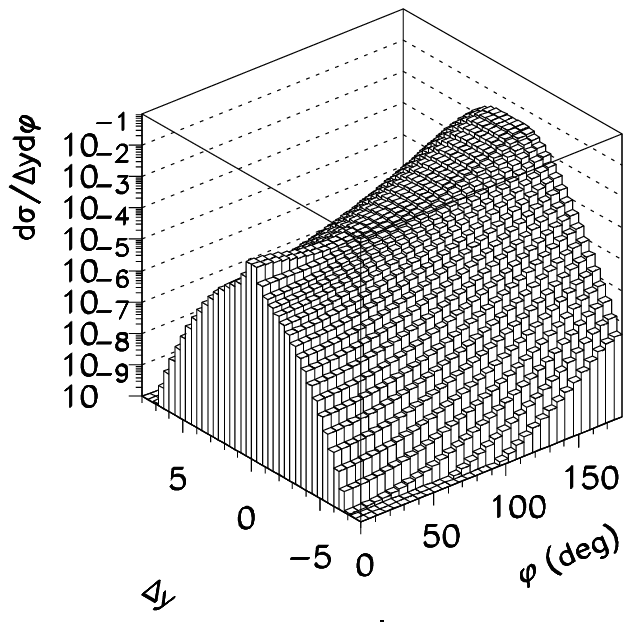}
\caption[*]{
$d\sigma/d\phi$ versus rapidity gap between $c$ and $\bar c$ for W = 1960 GeV$^2$.
($p_{1,t} = p_{2,t} \to (5, 15)$~GeV, $\Delta y = y_1 - y_2$, 
$\mu^2 = 10$~GeV$^2$  (left panel), $\mu^2 = 100$~GeV$^2$ (right panel).
\label{fig:dsig_dpt_gauss6}
}
\end{center}
\end{figure}

It was predicted for jet-jet correlations some time ago that a large rapidity
gap between jets leads to increased decorrelation in azimuth.
Can one expect a similar interesting effect for $c \bar c$ correlations?
In Fig.\ref{fig:dsig_dpt_gauss6} we show our result for $c \bar c$ correlations as a function of
the rapidity gap between $c$ and $\bar c$. Of course, the bigger rapidity distance 
between $c$ and $\bar c$ the smaller the cross section. However the shape
of the distribution in azimuth stays practically the same, except at
$\Delta y \approx$ 0, where a singularity for collinear quark and antiquark emissions
shows up.

\begin{figure}[!thb] % Figure 7
\begin{center}
\includegraphics[width=7cm]{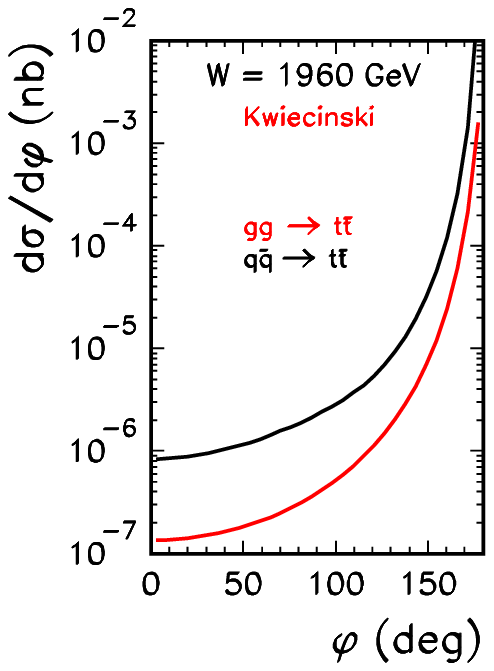}
\includegraphics[width=7cm]{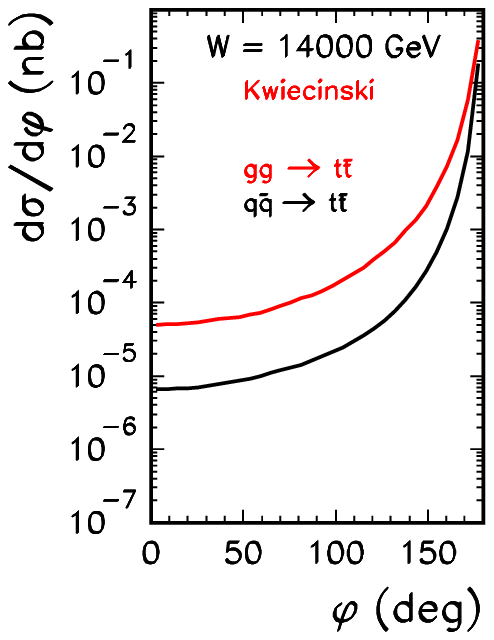}
\caption[*]{
(Color on-line) $d\sigma/d\phi$ for $t \bar t$ production at  W = 1960 GeV (left panel) 
and W = 14000 GeV (right panel) for two process: 
gg $\to t \bar t$ and $q \bar q \to t \bar t$ for Kwieci\'nski distributions. In this calculation
$\mu^2 = 10000$~GeV$^2$. 
\label{fig:dsig_dpt_gauss7}
}
\end{center}
\end{figure}

Finally we wish to come to azimuthal correlations of $t$ and $\bar t$.
Such an analysis seems possible even at present at the Tevatron.
In Fig.\ref{fig:dsig_dpt_gauss7} we show the azimuthal correlation function for two energies:
W = 1960 GeV (Tevatron) and W = 14 000 GeV (LHC). We show separately
two contributions: $g g \to t \bar t$ (red) and $q \bar q \to t \bar t$ (black).
While at the Tevatron energy the $q \bar q \to t \bar t$ contribution
dominates over the $g g \to t \bar t$ contribution, the situation reverses
at the LHC energy where the $g g \to t \bar t$ is the dominant contribution.
The shape of both components is, however, very similar.

It would be very interesting to compare the result obtained 
with in the $k_t$-factorization approach here with the result 
of the NLO collinear-factorization approach. The work in this direction is in progress.

%--------------------------------------------------------------------
%--------------------------------------------------------------------
%--------------------------------------------------------------------

%----------------------------
\section{Summary and outlook}
%----------------------------

Summarizing, we have discussed the (heavy quark) -- (heavy antiquark)
correlations in proton-proton and proton-antiproton collisions
within the $k_t$-factorization approach. Different unintegrated gluon
distributions have been used in the calculation. The results for azimuthal
angle as well as in $p_{1,t} \times p_{2,t}$ correlations have been presented.

In principle, corresponding experimental results would provide new information
and could test the unintegrated gluon distributions from the literature.
In practice, this could be more difficult as one measures rather heavy mesons
or charged leptons from the decays and only correlations of such objects
can be studied. However, both heavy mesons and charged leptons from the decays
"remember" to large extend the direction of initial heavy quark/antiquark.

On the theoretical side, a relation between $k_t$-factorization and standard
NLO approaches should be better understood and clarified.

We expect exciting future studies of $Q \bar Q$ correlations at the LHC,
especially for $t \bar t$ where detailed studies will be possible for the
first time in the history.

\vskip 0.5cm

{\bf Acknowledgments}
This work was partially supported by the grant of the Polish Ministry
of Scientific Research and Information Technology number 1 P03B 028 28.

%======================================================================

%-------------------------------
\end{document}